\documentclass[a4paper,10pt,twoside]{cpc-hepnp}
\usepackage{CJK,upgreek,fancyhdr}
\usepackage{subfigure}
\usepackage{multicol}
\usepackage{booktabs}
\usepackage{amsmath}
\usepackage{amsfonts,amssymb,bm,mathrsfs,bbm,amscd}
\usepackage{lastpage}
\usepackage[pdftex]{graphicx}
\usepackage{multirow}
\DeclareGraphicsExtensions{.pdf,.jpeg,.png}
\usepackage{epstopdf}
\usepackage[english]{babel}
\usepackage[colorlinks=false,pdfpagemode=FullScreen,,setpagesize=off,pdfborder={0 0 0}]{hyperref}
\usepackage{overpic}
\usepackage{lineno}
\usepackage{color}
\usepackage{rotating}
\usepackage{diagbox}
\usepackage{pdfsync}
\usepackage{hyperref}
\hypersetup{
	colorlinks=true,
	linkcolor=blue,
	filecolor=blue,
	urlcolor=blue,
	citecolor=blue,
}

\begin{document}
\begin{CJK*}{UTF8}{gkai}

\fancyhead[c]{\small Chinese Physics C~~~Vol. xx, No. x (2021) xxxxxx}
\fancyfoot[C]{\small 010201-\thepage}
\footnotetext[0]{Received xxxx June xxxx}

\title{Search for the weak decay $\psi(3686) \to \Lambda_c^{+} \bar{\Sigma}^- +c.c.$\thanks{
This work is supported in part by National Key Research and Development Program of China under Contracts Nos. 2020YFA0406400, 2020YFA0406300; National Natural Science Foundation of China (NSFC) under Contracts Nos. 11975118, 11635010, 11735014, 11835012, 11935015, 11935016, 11935018, 11961141012, 12022510, 12025502, 12035009, 12035013, 12192260, 12192261, 12192262, 12192263, 12192264, 12192265; 12061131003; the Natural Science Foundation of Hunan Province of China under Contract No. 2019JJ30019;  the Science and Technology Innovation Program of Hunan Province under Contract No. 2020RC3054; the Chinese Academy of Sciences (CAS) Large-Scale Scientific Facility Program; Joint Large-Scale Scientific Facility Funds of the NSFC and CAS under Contract No. U1832207; CAS Key Research Program of Frontier Sciences under Contract No. QYZDJ-SSW-SLH040; 100 Talents Program of CAS; The Institute of Nuclear and Particle Physics (INPAC) and Shanghai Key Laboratory for Particle Physics and Cosmology; ERC under Contract No. 758462; European Union's Horizon 2020 research and innovation programme under Marie Sklodowska-Curie grant agreement under Contract No. 894790; German Research Foundation DFG under Contracts Nos. 443159800, Collaborative Research Center CRC 1044, GRK 2149; Istituto Nazionale di Fisica Nucleare, Italy; Ministry of Development of Turkey under Contract No. DPT2006K-120470; National Science and Technology fund; National Science Research and Innovation Fund (NSRF) via the Program Management Unit for Human Resources and Institutional Development, Research and Innovation under Contract No. B16F640076; STFC (United Kingdom); Suranaree University of Technology (SUT), Thailand Science Research and Innovation (TSRI), and National Science Research and Innovation Fund (NSRF) under Contract No. 160355; The Royal Society, UK under Contracts Nos. DH140054, DH160214; The Swedish Research Council; U. S. Department of Energy under Contract No. DE-FG02-05ER41374.}}
\maketitle
\begin{center}
\begin{small}
\begin{center}
M.~Ablikim(麦迪娜)$^{1}$, M.~N.~Achasov$^{11,b}$, P.~Adlarson$^{70}$, M.~Albrecht$^{4}$, R.~Aliberti$^{31}$, A.~Amoroso$^{69A,69C}$, M.~R.~An(安美儒)$^{35}$, Q.~An(安琪)$^{53,66}$, X.~H.~Bai(白旭红)$^{61}$, Y.~Bai(白羽)$^{52}$
% ...
\\
\vspace{0.2cm}
(BESIII Collaboration)\\
\vspace{0.2cm} {\it
$^{1}$ Institute of High Energy Physics, Beijing 100049, People's Republic of China\\
% ...
$^{4}$ Bochum Ruhr-University, D-44780 Bochum, Germany\\
% ...
$^{11}$ G.I. Budker Institute of Nuclear Physics SB RAS (BINP), Novosibirsk 630090, Russia\\
% ...
$^{31}$ Johannes Gutenberg University of Mainz, Johann-Joachim-Becher-Weg 45, D-55099 Mainz, Germany\\
% ...
$^{35}$ Liaoning Normal University, Dalian 116029, People's Republic of China\\
% ...
$^{52}$ Southeast University, Nanjing 211100, People's Republic of China\\
$^{53}$ State Key Laboratory of Particle Detection and Electronics, Beijing 100049, Hefei 230026, People's Republic of China\\
% ...
$^{61}$ University of Jinan, Jinan 250022, People's Republic of China\\
% ...
$^{66}$ University of Science and Technology of China, Hefei 230026, People's Republic of China\\
% ...
$^{69}$ (A)University of Turin, I-10125, Turin, Italy; (B)University of Eastern Piedmont, I-15121, Alessandria, Italy; (C)INFN, I-10125, Turin, Italy\\
$^{70}$ Uppsala University, Box 516, SE-75120 Uppsala, Sweden\\
% ...
$^{b}$ Also at the Novosibirsk State University, Novosibirsk, 630090, Russia\\
% ...
}\end{center}

\vspace{0.4cm}
\end{small}
\end{center}

\begin{abstract}
Using $(448.1 \pm 2.9) \times 10^{6}$ $\psi(3686)$ events collected with the BESIII detector,
we perform the first
search for the weak baryonic decay $\psi(3686) \to \Lambda_c^{+} \bar{\Sigma}^- +c.c.$.
The analysis procedure is optimized using a blinded method.
No significant signal is observed, and the upper limit on the branching fraction ($\mathcal B$) of
$\psi(3686) \to \Lambda_c^{+} \bar{\Sigma}^- +c.c.$ is set as $1.4\times 10^{-5}$ at the 90\% confidence level.
\end{abstract}


\begin{multicols}{2}

\section{INTRODUCTION}

The weak decays of $J/\psi$ and $\psi(3686)$ are extremely rare compared to their dominant strong and electromagnetic decays.
For example, the branching fractions of the semi-leptonic and hadronic weak decays of the $J/\psi$
are predicted to be less than $10^{-9}$ in the framework of the standard model (SM)\cite{EPJC55.607, 2EPJC55.607}.
Over the past few years,
the BESIII collaboration previously searched for the baryon and lepton number violating decay $J/\psi \to \Lambda_c^+ e^-$\cite{prd99.072006} and the flavor changing neutral current decay $\psi(3686) \to \Lambda_c^+\bar p e^+e^-$\cite{prd97.091102}, as well as the weak decays $J/\psi \to D^{-}e^{+}\nu_{e}$\cite{BESIII:2021mnd} and
$J/\psi \to D_{s}^{(*)-}e^{+}\nu_{e}$\cite{BESIII:2014pps}. Throughout this paper, the charge conjugated channels are always implied.
To date, however, no signal has been observed in these channels.

Searches for purely baryonic weak $\psi(3686)$ decays involving a charmed baryon $\Lambda_c^+$ in the final state have not been previously performed.
Figure~\ref{fig:Feynman} shows the lowest order Feynman diagram for the rare baryonic decay  $\psi(3686) \to \Lambda_c^{+} \bar{\Sigma}^-$ in the SM. Here, $c$ quark acts as a spectator, the $\bar c$ quark
transforms into a $\bar{s}$, $d\bar{u}$ pair is produced via $W$-boson exchange, and $u \bar u$ pair is then produced from the vacuum.
These quarks and anti-quarks hadronize into $\Lambda^+_c$ and $\bar \Sigma^-$.
In Ref.~\cite{1102.5648}, the branching fraction of $\psi(3686) \to \Lambda_c^{+} \bar{\Sigma}^-$ was predicted to be of the order of $10^{-9}$ to $10^{-11}$ in the SM, which is comparable to those of the decays of $\psi(3686) \to$ charmed meson + anything.
New physics mechanisms beyond the SM, such as the top-color model\cite{Lu:2010rab} and Randall-Sundrum model\cite{Lillie:2007yh}, may enhance this decay branching fraction significantly.
The experimental study of $\psi(3686) \to \Lambda_c^{+} \bar{\Sigma}^-$ may therefore offer important information for a comprehensive understanding of the weak decay mechanisms of charmonium states.

A sample of $(448.1\pm2.9)\times 10^6$ $\psi(3686)$ events\cite{ref::psip-num-inc} is collected using electron and positron collisions, thereby offering an ideal opportunity to search for the $\psi(3686) \to \Lambda_{c}^{+}\bar{\Sigma}^{-}$ decay.
By analyzing this data sample, we report the first search for $\psi(3686) \to \Lambda_c^{+} \bar{\Sigma}^-$.

\begin{center}
	\centering
   \includegraphics[width=0.45\textwidth]{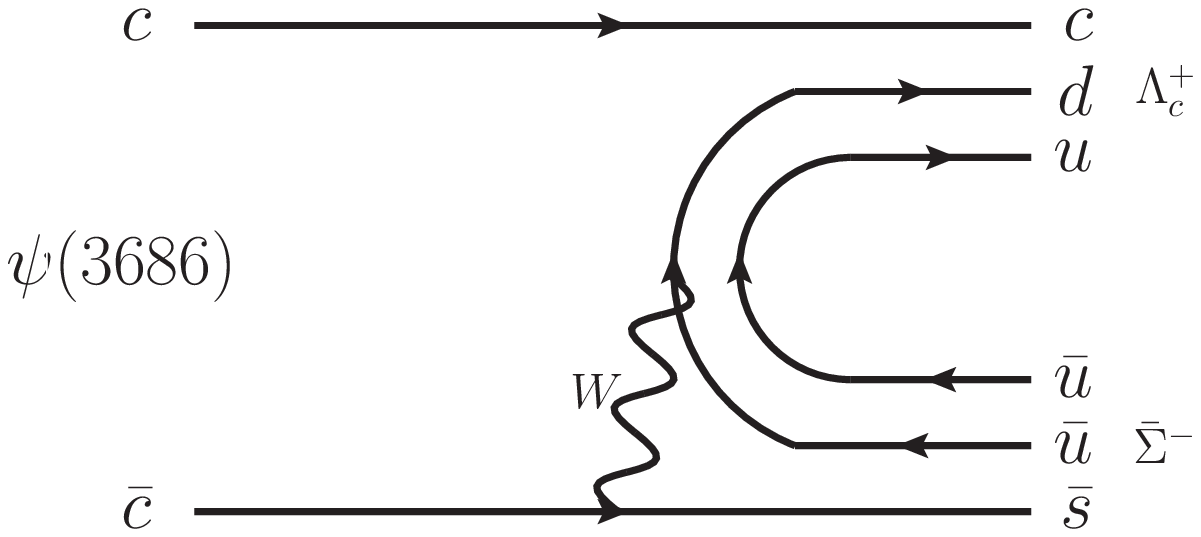}
	\figcaption{\label{Figure1} Feynman diagram for the process $\psi(3686) \to \Lambda_c^{+} \bar{\Sigma}^-$ in the SM\cite{1102.5648}.}
	\label{fig:Feynman}
\end{center}

\section{BESIII DETECTOR AND MONTE CARLO SIMULATION}

The BESIII detector\cite{Ablikim:2009aa} records symmetric $e^+e^-$ collisions
provided by the BEPCII storage ring\cite{Yu:IPAC2016-TUYA01}, which operates with a peak luminosity of $1\times10^{33}$~cm$^{-2}$s$^{-1}$
in the center-of-mass energy range from 2.0 to 4.95~GeV. BESIII has collected more than 32~fb$^{-1}$ of data samples in this energy region\cite{Ablikim:2019hff}. The cylindrical core of the BESIII detector consists of a helium-based
 multilayer drift chamber~(MDC), plastic scintillator time-of-flight
system~(TOF), and CsI(Tl) electromagnetic calorimeter~(EMC),
which are all enclosed in a superconducting solenoidal magnet
providing a 1.0~T magnetic field. The solenoid is supported by an
octagonal flux-return yoke with resistive plate counter muon
identification modules interleaved with steel. The charged-particle momentum resolution at $1~{\rm GeV}/c$ is
$0.5\%$, and the $dE/dx$ resolution is $6\%$ for electrons
from Bhabha scattering. The EMC measures photon energies with a
resolution of $2.5\%$ ($5\%$) at $1$~GeV in the barrel (end cap)
region. The time resolution in the TOF barrel region is 68~ps, while
that in the end cap region is 110~ps.

Simulated event samples produced with the {\sc geant4}-based\cite{geant4} Monte Carlo (MC) package, which
includes a geometric description of the BESIII detector and the
detector responses, are used to determine the detection efficiency
and estimate the backgrounds. The simulation includes the beam
energy spread and initial state radiation (ISR) in the $e^+e^-$
annihilations, modeled with the generator {\sc
kkmc}\cite{ref:kkmc,2ref:kkmc}.
The inclusive MC sample consists of the production of the charmonium resonances and the continuum processes incorporated in {\sc
kkmc}.
The known decay modes are modeled with {\sc
evtgen}\cite{ref:evtgen} using branching fractions taken from the
Particle Data Group\cite{ref::pdg2020}, and the remaining unknown decays
are taken from the charmonium states with {\sc
lundcharm}\cite{ref:lundcharm,2ref:lundcharm}. Final state radiation
from charged final-state particles is incorporated with the {\sc
photos} package\cite{photos}. In this analysis,
the $\psi(3686) \to \Lambda_{c}^{+}\bar{\Sigma}^{-}$
and $\bar{\Sigma}^{-} \to \bar{p}\pi^{0}$ are generated according to phase space,
and the $\Lambda_{c}^{+} \to pK^{-}\pi^{+}$ is generated using an amplitude analysis model\cite{L_c2pKpi}.
As in previous BESIII published papers\cite{prd99.072006,prd97.091102}, $\Lambda_c^+$ is only reconstructed via $\Lambda_{c}^{+} \to pK^{-}\pi^{+}$,
which provides the best sensitivity among all decay modes.

\section{EVENT SELECTION AND DATA ANALYSIS}

The procedure for selecting candidate events from the process $\psi(3686) \to \Lambda_c^{+} \bar{\Sigma}^-$,
where the $\Lambda_c^{+}$ baryon decays to $pK^{-}\pi^{+}$
and the $\bar{\Sigma}^-$ baryon decays to $\bar{p}\pi^{0}$,
is outlined below.

It is required that there are at least four charged tracks and two photons in the candidate events.
The polar angle of each charged track
is required to be in the range $|\rm{cos\theta}|<0.93$, coinciding with the coverage of the MDC.
The charged tracks from $\Lambda_{c}^{+} \to pK^{-}\pi^{+}$ decays must originate from the interaction
point with a distance of closest approach less than 1 cm on the transverse plane ($|V_{xy}|$)
and less than 10 cm along the $z$ axis ($|V_{z}|$).
For the charged tracks from $\bar{\Sigma}^-$ decays, the requirements of
$|V_{xy}|$ and $|V_{z}|$ are loosened to be less than 10 cm and 20 cm, respectively,
due to the relatively long lifetime of the $\bar{\Sigma}^-$.
No secondary vertex is considered because there is only one charged track from the $\bar{\Sigma}^-$.
Particle identification (PID) for the charged pion, kaon, and proton is performed using the $dE/dx$ and TOF information.
The particle type with the highest probability is assigned to each track.

 The $\pi^0$ candidates are identified as photon pairs reconstructed from the EMC showers. Each EMC shower is required to be within a 700 ns time window, which is applied to suppress electronic noise and energy depositions unrelated to the event. The energy deposited in nearby TOF counters is included in the energy of the EMC showers to improve the photon reconstruction efficiency and energy resolution.
 At least two photon candidates are required, with a minimum energy of 25~MeV in the barrel region ($|\rm cos\theta|<0.80$)
 or 50 MeV in the end cap region ($0.86<|\rm cos\theta|<0.92$).
The opening angle between the photon candidate and the nearest charged track must be greater than $10^{\circ}$.

A five-constraint (5C) kinematic fit is performed on the hypothesis of $e^+e^-\to  p K^{-} \pi^{+}  \bar{p} \gamma\gamma$,
with the invariant mass of the $\gamma\gamma$ combination constrained to the $\pi^0$ nominal mass.
The helix parameters of charged tracks of the MC events have been corrected to improve consistency with the data, following Ref. \cite{BESIII:2013nam}.
Events satisfying $\chi^2_{\rm 5C}<60$, which has been optimized based on the Punzi method \cite{punzi}, are kept for further analysis.
If there are multiple candidates in an event, the one with the smallest $\chi^2_{\rm 5C}$ is retained.

After applying all the above requirements, there are two main background sources \cite{top}, $\psi(3686) \to K^{*}(892)^{-} p\bar{\Lambda} $ ($K^{*}(892)^{-} \to \pi^{0} K^{-}$, $\bar{\Lambda} \to \pi^{+}\bar{p}$)
and $\psi(3686) \to \bar{K}^{*0}(892)p\bar{\Sigma}^{-} $ ($\bar{K}^{*0}(892) \to \pi^{+} K^{-}$, $\bar{\Sigma}^{-} \to \bar{p}\pi^{0}$). The former and latter are suppressed by requiring
the invariant mass of $\pi^{+}\bar{p}$ $(M(\pi^{+}\bar{p}))$ $\notin [1.090,1.130]$ GeV/$c^{2}$ and
the invariant mass of $K^{-}\pi^{+}$ $(M(K^{-}\pi^{+}))$ $\notin [0.756, 1.036]$ GeV/$c^{2}$, respectively.
These requirements have also been optimized with the Punzi method \cite{punzi}. The requirements of $M(\pi^+ \bar{p})$ and $M(K^-\pi^+)$ reject 94.2\% of background events with an efficiency loss of 45.5\%.
At present, the branching fraction of $\psi(3686) \to \bar{K}^{*0}(892)p\bar{\Sigma}^{-} $ is absent.
With the same $\psi(3686)$ data sample, we estimate the branching fraction of $\psi(3686) \to \bar{K}^{*0}(892)p\bar{\Sigma}^{-} +c.c.$
to be $(3.2  \pm 0.2_{\rm stat.})\times 10^{-3}$ by examining the $M(K^-\pi^+)$ distribution of the accepted candidate events.
Based on this, we estimate that the $\psi(3686) \to \bar{K}^{*0}(892)p\bar{\Sigma}^{-} $ background channel contributes to approximately 91 background events; however, they do not form a peak in
the $M(pK^{-}\pi^{+})$ distribution.

 Figure~\ref{fig:scatter}(a) and ~\ref{fig:scatter}(b) show the distributions of the invariant mass of $\bar{p}\pi^0$ ($M(\bar{p}\pi^0)$) versus the invariant mass of $pK^{-}\pi^{+}$ ($M(pK^{-}\pi^{+})$) surviving the event selection for the signal MC sample and data, respectively. The $\bar{\Sigma}^-$ candidates are required to be within the interval $M(\bar{p}\pi^0)\in (1.150, 1.230)$ GeV/$c^{2}$,
which corresponds to three times the resolution around $\bar \Sigma^-$ peak.
The signal yield of $\psi(3686) \to \Lambda_c^{+} \bar{\Sigma}^-$ is extracted from an unbinned maximum likelihood fit to the $M(pK^-\pi^+)$ distribution, as shown in Fig.~\ref{fig:fit_MKpi}.
In the fit, the lineshapes of signal and background are modeled by the signal
MC simulation and a first-order Chebyshev polynomial, respectively.
In addition, the yields of signal and background are free to float.
In Fig.~\ref{fig:fit_MKpi}, the background level is significantly higher than that expected by the inclusive $\psi(3686)$ MC sample,
which may be due to some unknown decay, for example, $\psi(3686) \to \pi^+K^-p \bar{\Sigma}^-+c.c.|_{{\rm non}-K^*}$.
With the same $\psi(3686)$ data sample, we estimate the branching fraction of $\psi(3686) \to \pi^+K^-p \bar{\Sigma}^-|_{{\rm non}-K^*}$
to be $(1.49\pm 0.03_{\rm stat.})\times 10^{-4}$ by examining the $M(p\pi^+)$ distribution of the accepted candidate events.
Based on this, we estimate that the $\psi(3686) \to \pi^+K^-p \bar{\Sigma}^-|_{{\rm non}-K^*}$ background channel contributes to approximately five background events; however, they do not form a peak in
the $M(pK^{-}\pi^{+})$ distribution. Although there is still room for various background channels, they do not all form a peak in the $M(pK^{-}\pi^{+})$ distribution.
Because no significant signal is observed from the $\psi(3686)$ data, conservative upper limits will assume that all the fitted signals are from $\bar{\Sigma}^-$ after the following two checks. First, the events in the $\bar{\Sigma}^-$ sideband region shows that the non-$\bar{\Sigma}^-$ contribution in the selected candidates is negligible.
Second, an analysis of 2.93 fb$^{-1}$ of data taken at $\sqrt s=$3.773 GeV shows that no peaking background of the continuum production of $e^+e^-\to \Lambda_c^{+} \bar{\Sigma}^-$ is estimated.

To estimate the upper limit on the branching fraction of $\psi(3686) \to \Lambda_c^{+} \bar{\Sigma}^-$, we use a likelihood scan method after incorporating systematic uncertainties as discussed in next section.

\begin{center}
    \includegraphics[width=0.45\textwidth]{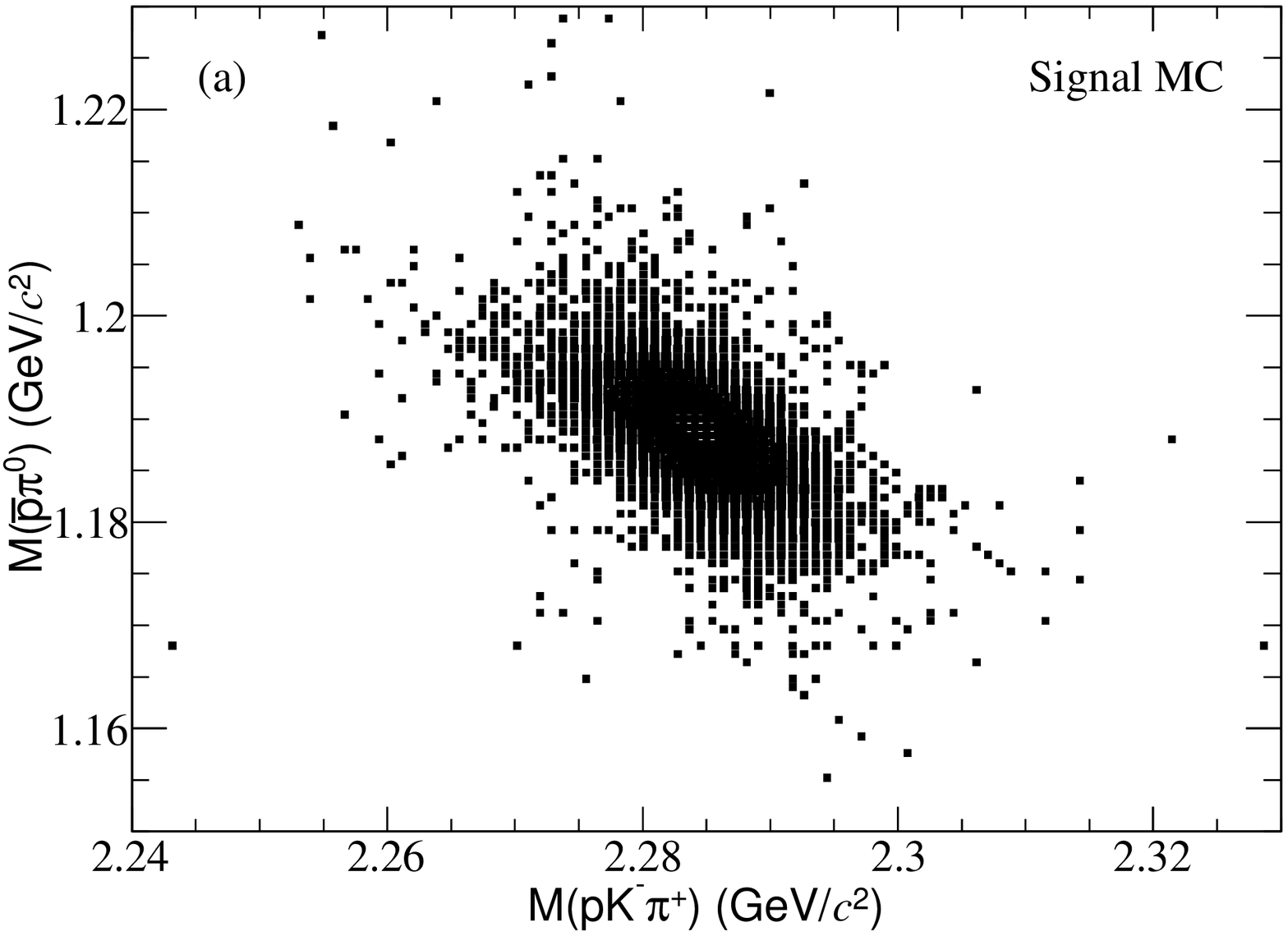}
    \includegraphics[width=0.45\textwidth]{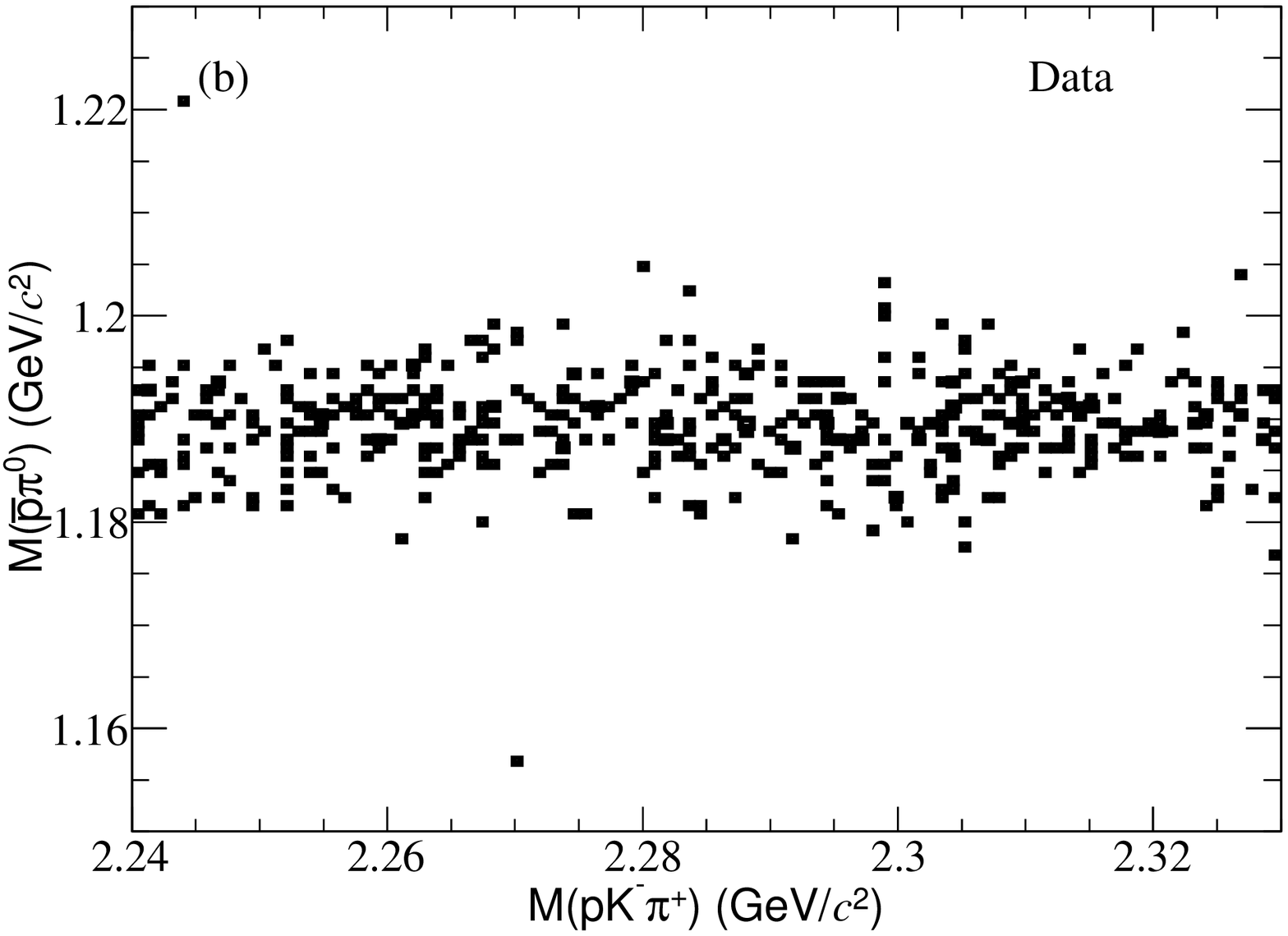}
	\figcaption{ \label{Figure2} Distribution of $M(\bar{p}\pi^0)$ versus $M(pK^{-}\pi^{+})$ for the accepted candidate events in the signal MC sample and data.}
    \label{fig:scatter}
\end{center}

\begin{center}
\includegraphics[width=0.45\textwidth]{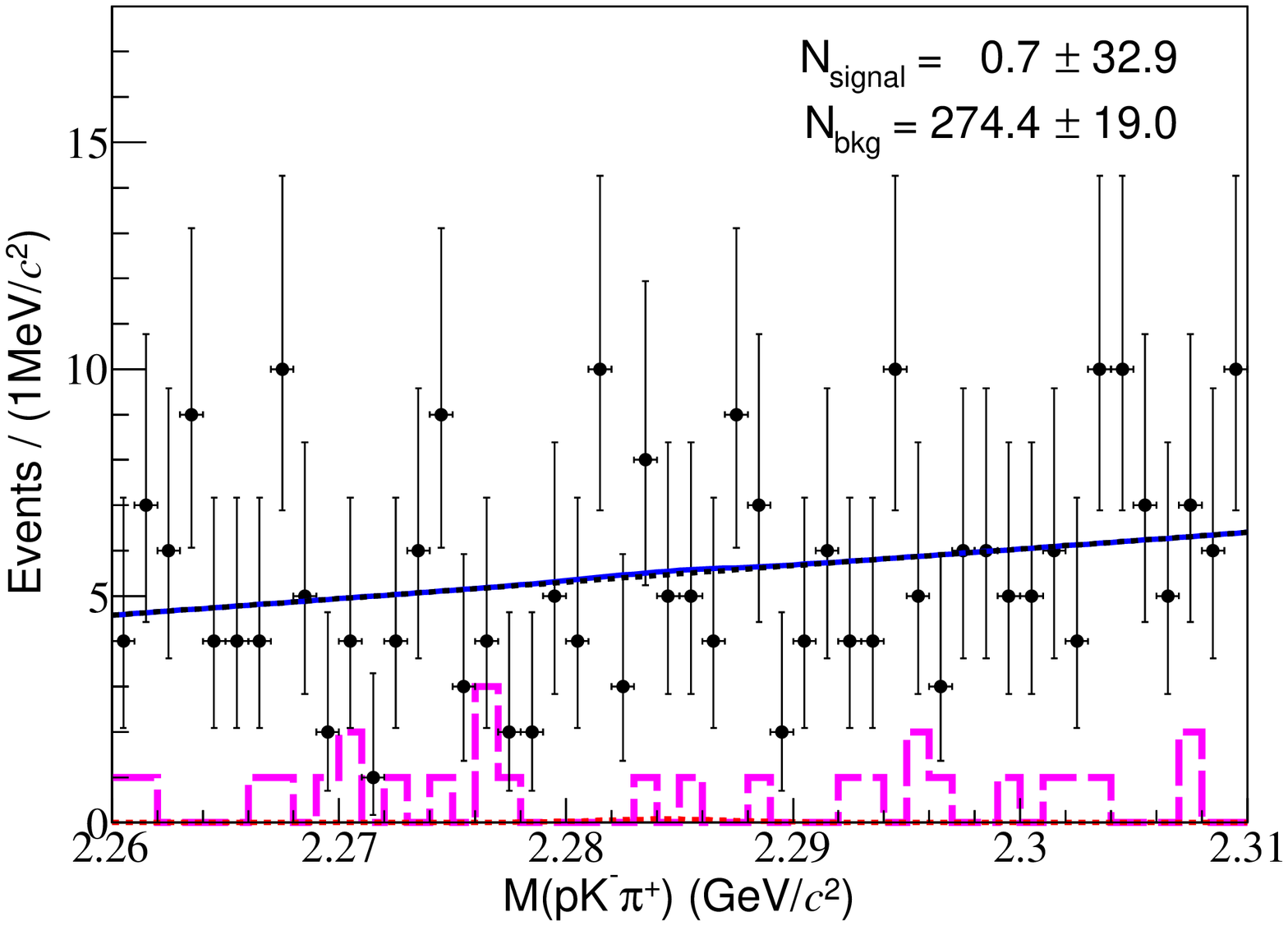}
\figcaption{\label{Figure3} (color online)Fit to the $M(pK^-\pi^+)$ distribution for the candidate events from $\psi(3686) \to \Lambda_c^{+} \bar{\Sigma}^-$. Points with error bars are data. The red (black) dashed line is the signal (background), and the blue solid curve is the total fit. The pink dashed line is the inclusive MC sample. The red solid curve is the signal shape enlarged by a factor of 100.}
\label{fig:fit_MKpi}
\end{center}

\section{SYSTEMATIC UNCETTAINTY}

Systematic uncertainties for the upper limit on the branching fraction  can be classified into two categories: additive terms and multiplicative terms.

Additive terms contain uncertainties caused by the chosen signal shape, background shape, and fit range. The effect due to the signal shape is estimated by replacing the signal MC shape with the signal MC shape convolved with a Gaussian resolution function with a mean of 1.0 MeV/$c^2$ and a resolution of 1.3 MeV/$c^2$. These parameters are obtained from a fit to the $M(pK^{-}\pi^{+})$ spectrum using a data sample taken above $\Lambda^+_c\bar \Lambda^-_c$ production threshold. The effect from the background shape is evaluated using first-order and second-order Chebyshev polynomials.
The effect from the fit range is estimated with fit ranges of $[2.26,2,31]$ GeV/$c^{2}$, $[2.26,2,32]$ GeV/$c^{2}$ and $[2.25,2,31]$ GeV/$c^{2}$.
Among all of these terms, the case yielding the largest upper limit is chosen for further analysis.

The sources of multiplicative systematic uncertainties include the number of $\psi(3686)$ events, tracking efficiency, PID efficiency, ${\pi^{0}}$ reconstruction, the $\bar{\Sigma}^{-}$ mass window, kinematic fit, the quoted branching fractions of intermediate states, and the signal MC model. The systematic uncertainties on the requirements of $M(\pi^{+} \bar{p})$ and $M(\pi^{+} K^{-})$ are estimated by changing individual veto regions by 10 MeV/$c^{2}$. The associated effects on the upper limits are less than 0.1\% which are negligible. The other systematic uncertainties are discussed below.

\begin{itemize}
\item[\bf (a)] {Number of $\psi(3686)$ events}: The total number of $ \psi(3686)$ events in the data sample was determined to be $(448.1 \pm 2.9) \times 10^{6}$ with inclusive hadronic events in Ref.\cite{ref::psip-num-inc}. The uncertainty of the total number of $\psi(3686)$ events, $0.6$\%, is assigned as a systematic uncertainty.

\item[\bf (b)] {Tracking and PID efficiencies:} The uncertainties from the tracking and PID efficiencies have been studied with the high purity control samples $\psi(3686) \to \pi^{+}\pi^{-}J/\psi$\cite{pikif}. The systematic uncertainty due to the tracking or PID efficiency is assigned to be 1.0\% for each track.

\item[\bf (c)] {${\pi^{0}}$ reconstruction:} The systematic uncertainty of the $\pi^{0}$ reconstruction efficiency has been studied with the control sample of $J/\psi \to \rho\pi$ in Ref.\cite{pikif}. The associated systematic uncertainty is assigned to be 1.0\% for each $\pi^{0}$.

\item[\bf (d)] {$\bar{\Sigma}^-$ mass window:} To estimate the systematic uncertainty from the $\bar{\Sigma}^-$ mass window, we use the control sample of $\psi(3686) \rightarrow \Sigma^{+} \bar{\Sigma}^-$ with $\Sigma^{+} \rightarrow p \pi^0$ and $\bar{\Sigma}^- \rightarrow \bar{p} \pi^0$. The difference between the acceptance efficiencies of data and MC simulation, 0.1\%, is taken as the corresponding systematic uncertainty.

\item[\bf (e)] {5C kinematic fit:} To examine the systematic uncertainty due to the 5C kinematic fit, we examine the signal efficiencies with and without correcting the MDC helix parameters for the signal MC events. The change in the signal efficiency, 0.2\%, is assigned as the systematic uncertainty.

\item[\bf (e)] {Quoted branching fraction:} The branching fractions of $\Lambda_c^+\to pK^-\pi^+$, $\bar{\Sigma^-}\to\bar{p}\pi^0$, and $\pi^0\to\gamma\gamma$ are quoted from the Particle Data Group\cite{ref::pdg2020}, which are $(6.28 \pm 0.32)\%$, $(51.57 \pm 0.30)\%$, and $(98.823 \pm 0.034)\%$, respectively. They contribute to a total uncertainty of 5.2\%, which is regarded as a systematic uncertainty.

\item[\bf (f)] {MC model:}
 The signal MC sample of $\psi(3686) \to \Lambda_{c}^{+} \bar{\Sigma}^-$ is generated according to phase space. To estimate the systematic uncertainty on the MC model, we generate alternative signal MC samples using the J2BB1 model\cite{Ping:2008zz} with an angular distribution of $1+\alpha\cos^{2}\theta$.
To be conservative, two extreme scenarios corresponding to $\alpha = -1$ and $\alpha = 1$ are considered. The difference in the efficiencies between the phase space model and the J2BB1 model, 11.0\%, is taken as the corresponding systematic uncertainty.
 \end{itemize}

Assuming that all sources are independent, the total multiplicative systematic uncertainty is determined to be 13.5\% by adding all uncertainties quadratically.
The systematic uncertainties are summarized in Table \ref{tab:Sys}.

\begin{center}
\tabcaption{ Multiplicative systematic uncertainties in the branching fraction measurement.}
\label{tab:Sys}
\footnotesize
\begin{tabular}{lc}\hline
  Source                        & Uncertainty (\%)   \\ \hline
  Number of $\psi(3686)$ events & 0.6  \\
  Tracking efficiencies         & 4.0  \\
  PID  efficiencies             & 4.0  \\
  $\pi^{0}$ reconstruction      & 1.0  \\
  $\bar{\Sigma}^-$ mass window  & 0.1  \\
  5C kinematic fit              & 0.2  \\
  Quoted branching fractions    & 5.2  \\
  MC model                      & 11.0  \\
  Total                         & 13.5  \\
\hline
\end{tabular}
\end{center}

\begin{center}
\centering
\includegraphics[width=0.45\textwidth]{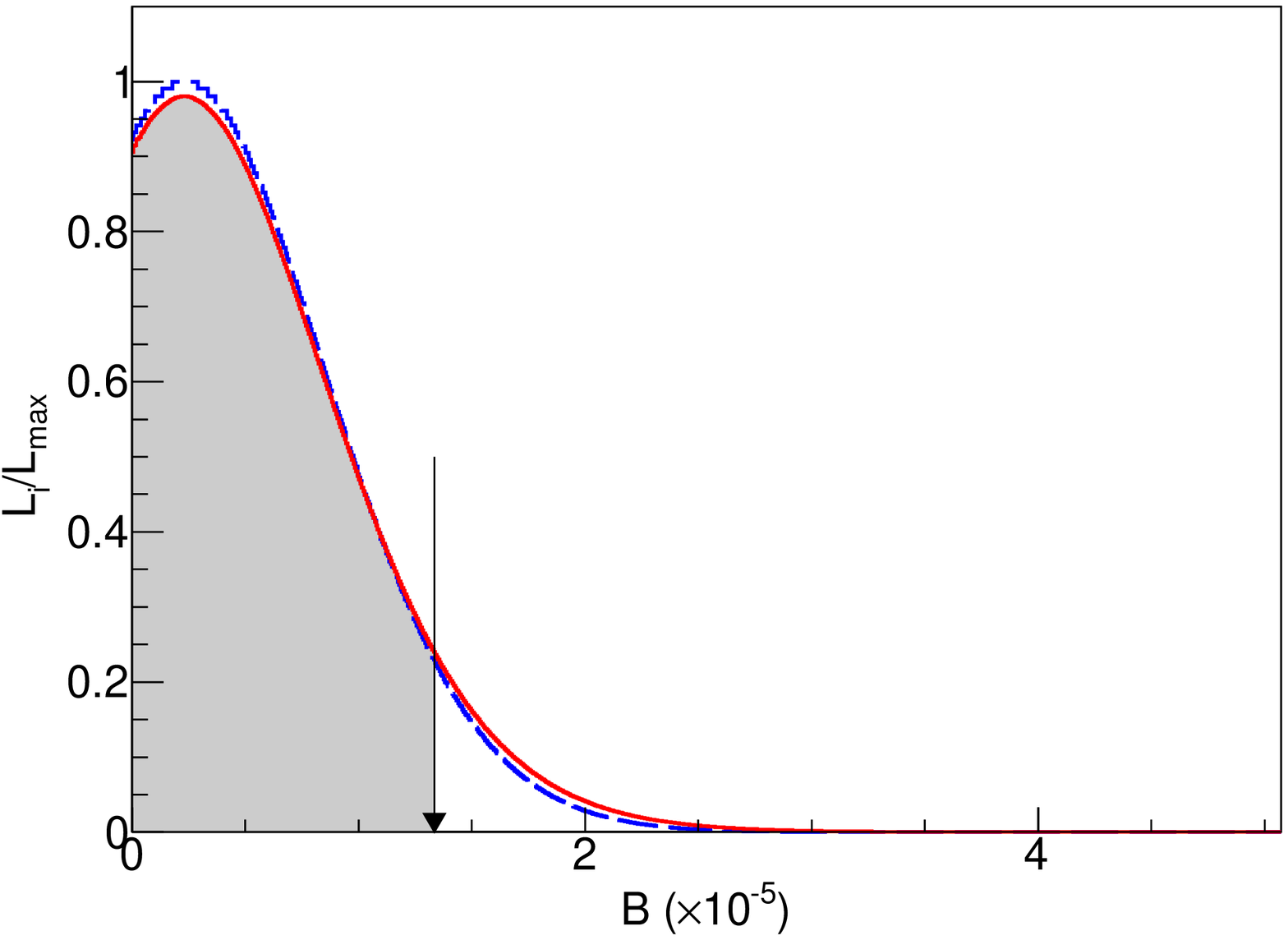}
\figcaption{\label{Figure4}(color online) Distributions of the likelihoods versus the branching fraction of $\psi(3686) \to \Lambda_c^{+} \bar{\Sigma}^-$.
The results obtained with and without incorporating the systematic uncertainties are shown in the red solid and blue dashed curves, respectively. The black arrow shows the result corresponding to the 90\% confidence level.}
\label{fig:prob}
\end{center}

\section{RESULT}

The branching fraction of $\psi(3686) \to \Lambda_c^{+} \bar{\Sigma}^-$ is calculated using
\begin{equation} \label{eq:sigmaobs}
\mathcal{ B}(\psi(3686) \to \Lambda_c^{+} \bar{\Sigma}^-) = \frac{ N_{\rm sig}}{ N_{\rm \psi(3686)} \cdot {\Pi {\mathcal{ B}_{i}}} \cdot {\epsilon}},
\end{equation}
where $N_{\psi(3686)} $ is the total number of $\psi(3686)$ events in the data sample, $\Pi \mathcal{ B}_i$ is the
product of the branching fractions of the intermediate decays $\Lambda_c^{+} \to p K^{-} \pi^{+}$,
$\bar{\Sigma}^- \to \bar{p} \pi^{0}$, and $\pi^{0} \to \gamma \gamma$,
and ${\epsilon}$ is the detection efficiency, which is determined to be $(11.03\pm0.08)\%$ based on MC simulation.

No significant signal is observed, and the upper limit on the signal yield is set to be 21.1 at the 90\% confidence level by assuming the fitted signal yield is entirely from the process $\psi(3686) \to \Lambda_c^{+} \bar{\Sigma}^-$.
The raw likelihood distribution versus ${\mathcal B}(\psi(3686) \to \Lambda_c^{+} \bar{\Sigma}^-)$ is represented by
the blue dashed curve in Fig.~\ref{fig:prob}.
This curve is then smeared by a Gaussian function with a mean of 0 and a width equal to the multiplicative systematic uncertainty of $13.5\%$ according to Refs.\cite{K.Stenson:2006,cpc:up}.
The updated likelihood distribution is shown as the red solid curve in Fig.~\ref{fig:prob}.
By integrating the red dashed curve from zero to 90\% of the physical region,
the upper limit on the branching fraction of $\psi(3686) \to \Lambda_c^{+} \bar{\Sigma}^-$ at the 90\% confidence level is set to be
\begin{eqnarray}
{\mathcal B}(\psi(3686) \to \Lambda_c^{+} \bar{\Sigma}^-)< 1.4\times 10^{-5}. \nonumber
\end{eqnarray}

\section{SUMMARY}

By analyzing $(448.1 \pm 2.9) \times 10^{6}$ $\psi(3686)$ events collected with the BESIII detector,
we present the first search for $\psi(3686) \to \Lambda_c^{+} \bar{\Sigma}^-$.
No significant signal is observed in the data sample.
Therefore, we set an upper limit on the branching fraction of $1.4\times 10^{-5}$ at the 90\% confidence level.
This is far above the prediction in the SM. An additional 2.3 billion of $\psi(3686)$ events at BESIII will be available soon.
This larger $\psi(3686)$ data sample offers an opportunity to further improve the sensitivity of the search for this decay\cite{Ablikim:2019hff}.

\acknowledgments{
The BESIII collaboration thanks the staff of BEPCII and the IHEP computing center for their strong support.
}

\end{multicols}

\vspace{-1mm}
\centerline{\rule{80mm}{0.1pt}}
\vspace{2mm}

\begin{multicols}{2}

\end{multicols}

\clearpage
\end{CJK*}
\end{document}